\documentclass{ws-ijgmmp}

\usepackage{amssymb}
\usepackage{amsmath}

\def\beq{\begin{equation}}
\def\eeq{\end{equation}}
\def\bea{\begin{align}}
\def\eea{\end{align}}
\def\beqn{\begin{equation*}}
\def\eeqn{\end{equation*}}

\def\aa{\alpha}
\def\bb{\beta}
\def\cc{\gamma}
\def\dd{\delta}
\def\ee{\epsilon}

\def\nn{\eta}
\def\oo{\omega}

\def\CC{\Gamma}

\def\bk{\boldsymbol{k}}
\def\be{\boldsymbol{e}}

\def\Ppmm<#1,#2>{{\hbox{${\lower5pt\hbox{$#2$}}
\atop{\smash{\lower1.8pt\hbox{$\scriptscriptstyle #1$}}}$}}}
\def\G<#1>{\Ppmm<#1,\Gamma>}
\def\GG<#1>{\Ppmm<#1,G>}
\def\R<#1>{\Ppmm<#1,R>}
\def\T<#1>{\Ppmm<#1,T>}

\def\mD{\mathcal D}

\def\rmd{{\rm d}}
\begin{document}

\markboth{Donato Bini, Andrea Geralico and Roy P. Kerr}
{The Kerr-Schild ansatz revised}

\catchline{}{}{}{}{}
%
%%%%%%%%%%%%%%%%%%%%%%%%%%%%%%%%%%%%%%%%%%%%%%%%%%%%%%%%%%%%%%%%%%%%

\title{THE KERR-SCHILD ANSATZ REVISED}

\author{DONATO BINI}

\address{Istituto per le Applicazioni del Calcolo ``M. Picone,'' CNR I-00161 Rome, Italy\\ 
ICRA, University of Rome ``La Sapienza,'' I-00185 Rome, Italy\\
ICRANet, I-65100 Pescara, Italy\\
\email{binid@icra.it}}

\author{ANDREA GERALICO}

\address{Physics Department, University of Rome ``La Sapienza,'' I-00185 Rome, Italy\\ 
ICRA, University of Rome ``La Sapienza,'' I-00185 Rome, Italy\\
ICRANet, I-65100 Pescara, Italy\\
\email{geralico@icra.it}}

\author{ROY P. KERR}

\address{University of Canterbury, Christchurch, New Zealand\\ 
ICRANet, I-65100 Pescara, Italy\\
\email{roy.kerr@canterbury.ac.nz}}

\maketitle

\begin{history}
\received{(Day Month Year)}
\revised{(Day Month Year)}
\end{history}

\begin{abstract}
Kerr-Schild metrics have been introduced as a linear superposition of the flat spacetime metric and a squared null vector field, say $\bk$, multiplied by some scalar function, say $H$.
The basic assumption which led to Kerr solution was that $\bk$ be both geodesic and shearfree.
This condition is relaxed here and Kerr-Schild ansatz is revised by treating Kerr-Schild metrics as {\it exact linear perturbations} of Minkowski spacetime.
The scalar function $H$ is taken as the perturbing function, so that Einstein's field equations are solved order by order in powers of $H$.
It turns out that the congruence must be geodesic and shearfree as a consequence of third and second order equations, leading to an alternative derivation of Kerr solution.
\end{abstract}


\section{Introduction}

Kerr-Schild metrics have the form \cite{kerrschild1,kerrschild2}
\beq
\label{KSmetric1}
\rmd s^2 = g_{\alpha\beta}\rmd x^\alpha\rmd x^\beta\equiv (\eta_{\alpha\beta}-2Hk_\alpha k_\beta)\rmd x^\alpha\rmd x^\beta\ ,
\eeq
where $\eta_{\alpha\beta}$ is the metric for Minkowski space and $k_\aa$ is a null vector
\beq
 \label{KSmetric2}
    \eta_{\alpha\beta}k^\aa k^\bb =g_{\alpha\beta}k^\aa k^\bb = 0,\qquad k^\aa = \eta^{\alpha\beta}k_\bb= g^{\alpha\beta}k_\bb\ .
\eeq
The inverse metric is also linear in $H$
\beq
\label{KSmetric3}
    g^{\aa\bb} = \eta^{\aa\bb} + 2Hk^\alpha k^\beta\ ,
\eeq
and so the determinant of the metric is independent of $H$
\beq
    (\eta_{\alpha\cc}-2Hk_\alpha k_\cc)(\eta^{\cc\bb} + 2Hk^\cc k^\bb)= \delta_\aa^\bb \quad \longrightarrow\quad |g_{\alpha\beta}| = |\eta_{\alpha\beta}|\ .
\eeq

Within this class of general metrics the Kerr solution was obtained in 1963 by a systematic study of algebraically special vacuum solutions \cite{kerr1}.
If $(x^0=t,x^1=x,x^2=y,x^3=z)$ are the standard Cartesian coordinates for Minkowski spacetime with $\eta_{\alpha\beta}={\rm diag}[-1,1,1,1]$, then for Kerr metric we have
\begin{equation}
\label{KSmetric4}
-k_\alpha\rmd x^\alpha=\rmd t +\frac{(rx+ay)\rmd x+(ry-ax)\rmd y}{r^2+a^2} + \frac{z}{r}\rmd z\ ,
\end{equation}
where $r$ and $H$ are defined implicitly by
\begin{equation}
\label{KSmetric5}
\frac{x^2+y^2}{r^2+a^2}+\frac{z^2}{r^2}=1\ ,\qquad
H=-\frac{{\mathcal M}r^3}{r^4+a^2z^2}\ .
\end{equation}

Kerr solution is asymptotically flat and the constants ${\mathcal M}$ and $a$ are the total mass and specific angular momentum for a localized source. They both have the dimension of a length in geometrized units.
The vector $\bk$ is geodesic and shearfree, implying that Kerr metric is algebraically special according to the Goldberg-Sachs theorem \cite{goldberg}.
Moreover, $\bk$ is independent of ${\mathcal M}$ and hence a function of $a$ alone.
Note that the mass parameter ${\mathcal M}$ appears linearly in the metric, i.e. in $H$.

In this paper we consider Kerr-Schild metrics (\ref{KSmetric1}) as {\it exact linear perturbations} of Minkowski space and solve Einstein's field equations order by order in powers of $H$.
The results of this analysis will be that $\bk$ must be geodesic and shearfree as a consequence of third and second order equations, leading to an alternative derivation of Kerr solution.

\section{Modified ansatz}

Let $\ee$ be an arbitrary constant  parameter, eventually to be set equal to $1$, so that the Kerr-Schild metric (\ref{KSmetric1}) reads
\beq\label{ansatz}
    g_{\aa\bb} = \eta_{\aa\bb} - 2 \ee H k_\aa k_\bb\ ,
\eeq
with inverse
\beq
    g^{\aa\bb} = \eta^{\aa\bb} + 2 \ee H k^\aa k^\bb\ ,
\eeq
and suppose that coordinates are chosen so that the components  $\nn_{\aa\bb}$ are constants, but not necessarily of the form $\eta_{\alpha\beta}={\rm diag}[-1,1,1,1]$.
The connection is then quadratic in $\ee$
\beq
\CC^\cc{}_{\aa\bb} = \ee\G<1>^\cc{}_{\aa\bb} + \ee^2\G<2>^\cc{}_{\aa\bb}\ ,
\eeq
where
\begin{align}
    \G<1>^\cc{}_{\aa\bb} &=- (Hk_\aa k^\cc )_{,\bb} - (Hk_\bb k^\cc)_{,\aa}+(Hk_\aa k_\bb)_{,\lambda}\eta^{\lambda\cc}\ ,\nonumber\\
    \G<2>^\cc{}_{\aa\bb} &=2H[H(\dot k_\aa k_\bb + \dot k_\bb k_\aa)+\dot H k_\aa k_\bb]k^\cc\equiv2Hk^\cc(Hk_\aa k^\bb)\!\!{\dot{\phantom{X}}}\ ,
\end{align}
a ``dot'' denoting differentiation in the $\bk$ direction, i.e. $\dot f=\bk(f)=f_{,\aa}k^{\aa}$.
Note that only the indices of $\bk$ can be raised and lowered with the Minkowski metric.
Hereafter we will use an ``index''  $0$ to denote contraction with $\bk$, i.e.
\begin{align}
    \CC^0{}_{\aa\bb}  &= \CC^\cc{}_{\aa\bb} k_\cc = \ \ \ee(H k_\aa k_\bb)\!\!{\dot{\phantom{X}}}\ ,\nonumber\\
    \CC^\cc{}_{\aa 0} &= \CC^\cc{}_{\aa\bb} k^\bb = - \ee(H k_\aa k^\cc)\!\!{\dot{\phantom{X}}}\ ,\nonumber\\
    \CC^\cc{}_{0 0} &= \CC^\cc{}_{\aa\bb} k ^\aa k^\bb = 0\ ,\nonumber\\
    \CC^0{}_{\aa 0} &= \CC^\cc{}_{\aa\bb} k ^\bb k_\cc = 0\ .
\end{align}

The determinant of the full metric is independent of $\ee$
\beq
    |g_{\aa\bb}| = |\eta_{\aa\bb} - 2 \ee H k_\aa k_\bb| = |\eta_{\aa\bb}|={\rm const.}
    \qquad\longrightarrow \qquad \CC^\bb{}_{\aa\bb} = 0\ ,
\eeq
and the contracted Riemann tensor therefore reduces to
\beq
    R_{\aa\bb} = {R^\cc}_{\aa\cc\bb} = \CC^\cc{}_{\aa\bb,\cc}
    - \CC^\cc{}_{\aa\dd} \CC^\dd{}_{\bb\cc}\ .
\eeq
The simplest component is
\beq\begin{split}
   R_{\aa\bb} k^\aa k^\bb &= \CC^\cc{}_{\aa\bb,\cc} k^\aa k^\bb
    - \CC^\cc{}_{\dd 0} \CC^\dd{}_{\cc 0} = \CC^\cc{}_{00,\cc} - 2 \CC^\cc{}_{\aa 0} k^\aa{}_{,\cc}\\
    &= 2 \ee H||\dot \bk||^2\ .
\end{split}\eeq

In vacuum the LHS is zero, then $||\dot \bk|| = 0$ and so $\dot
\bk$ is a null-vector orthogonal to another null-vector, $\bk$.
Hence $\dot \bk$ must be parallel to $\bk$ and therefore $\bk$ is
a geodesic vector.

The Ricci tensor expanded as series in $\ee$ is given by
\beq
   R_{\aa\bb} = \ee\R<1>_{\aa\bb} + \ee^2\R<2>_{\aa\bb} + \ee^3\R<3>_{\aa\bb} + \ee^4\R<4>_{\aa\bb}\ .
\eeq
The vacuum Einstein's equations $R_{\aa\bb} = 0$ imply that contributions of all orders must vanish.
Let us evaluate all such components.

The highest components of the expansion for the Ricci tensor are
\begin{align}
\label{ricci4}
    \R<4>_{\aa\bb} =& -\Ppmm<2, \CC>^\rho{}_{\aa\sigma}
    \Ppmm<2, \CC>^\sigma{}_{\bb\rho}= 0\ ,\\
\label{ricci3}
    \R<3>_{\aa\bb} =& -\G<1>^\rho{}_{\aa\sigma}
    \Ppmm<2, \CC>^\sigma{}_{\bb\rho} -\Ppmm<2, \CC>^\rho{}_{\aa\sigma}
    \G<1>^\sigma{}_{\bb\rho} = 4H^3||\dot \bk||^2k_\aa k_\bb\ .
\end{align}

The next component of $R_{\aa\bb}$ is
\beq
\label{ricci2}
\begin{split}
    \R<2>_{\aa\bb} &= \Ppmm<2, \CC>^\rho{}_{\aa\bb,\rho} -\G<1>^\rho{}_{\aa\sigma}  \G<1>^\sigma{}_{\bb\rho}\\
    & = 2H\left[(Hk_\aa k_\bb)\!\!{\ddot{\phantom{X}}} +k^\sigma{}_{,\sigma}(H k_\aa k_\bb)\!\!{\dot{\phantom{X}}}-H\dot k_\aa \dot k_\bb\right]\\
    &\qquad -H^2\Phi k_\aa k_\bb-2Hk_{(\aa} \psi_{\bb)}\ ,\\
\end{split}
\eeq
where
\beq
\Phi=4\eta^{\gamma\lambda}\eta^{\delta\mu}k_{[\lambda,\delta]} k_{[\mu,\gamma]}\ , \qquad
\psi_\alpha=2{\dot k}^{\gamma}(Hk_\aa)_{,\gamma}\ .
\eeq
Finally, the first component of the Ricci tensor expansion is
\beq
\label{ricci1}
\begin{split}
    \R<1>_{\aa\bb}&= \Ppmm<1, \CC>^\cc{}_{\aa\bb,\cc}\\
&= Ak_\aa k_\bb+2k_{(\aa}B_{\bb)}+X_{\aa\bb}\ ,\\
\end{split}
\eeq
where
\begin{align}
\label{ricci1_coeffs}
A&= \eta^{\lambda\gamma}H_{,\lambda\gamma}\ ,\nonumber\\
B_\bb &= -(Hk^\gamma)_{,\gamma\beta}+\frac{1}{H}\eta^{\lambda\gamma}(H^2k_{\beta , \gamma})_{,\lambda}\ ,\nonumber\\
X_{\aa\bb}&= -2H\left[(k_{(\alpha,\beta)}k^\gamma)_{,\gamma}+k_{(\alpha , |\gamma |}k^\gamma{}_{,\beta)}-\eta^{\lambda\gamma}k_{\alpha , \gamma}k_{\beta , \lambda}  \right]\nonumber \\
&\quad\, -2k^\gamma \left[H_{,(\alpha}k_{\beta ),\gamma}+H_{, \gamma}k_{(\alpha , \beta)}\right]\nonumber \\
&=-2H\left[\dot k_{(\alpha,\beta)}+k^\gamma{}_{,\gamma}k_{(\alpha,\beta)}-\eta^{\lambda\gamma}k_{\alpha , \gamma}k_{\beta , \lambda}\right]\nonumber \\
&\quad\,-2{\dot H}k_{(\alpha,\beta)}-2H_{,(\aa} \dot k_{\bb)}\ .
\end{align}

\subsection{Kinematical properties of the congruence $\bk$}

Taking the covariant derivative of $\bk$ gives
\beq
\nabla_\alpha
k_\beta=k_{\beta , \alpha}-\epsilon (Hk_\alpha
k_\beta)\!\!{\dot{\phantom{X}}}\ ,
\eeq
so that its
$4$-acceleration is simply
\beq
a(k)_\beta=k^\mu \nabla_\mu
k_\beta=\dot k_{\beta}\ .
\eeq The other optical scalars of
interest are the expansion
\beq
\label{th_def}
\theta=\frac12
k^\alpha{}_{;\alpha}=\frac12 \eta^{\alpha\beta}k_{\beta ,
\alpha}=\frac12 k^{\alpha}{}_{,\alpha}\ ,
\eeq
the vorticity
\beq
\label{om_def}
\omega^2=\frac12k_{[\alpha ; \beta]}k^{\alpha ;
\beta}=\frac12 k_{[\beta ,
\alpha]}\left(\eta^{\alpha\mu}\eta^{\beta\nu}k_{\mu,
\nu}-2\epsilon H \dot k^\alpha k^\beta \right)\ ,
\eeq
and the
shear, implicitly defined by
\beq
\label{shear_def}
\theta^2+|\sigma|^2=\frac12 k_{(\alpha ; \beta)}k^{\alpha ;
\beta}=\frac12 k_{(\beta ,
\alpha)}\eta^{\alpha\mu}\eta^{\beta\nu}k_{\mu, \nu}-\frac12
\epsilon H ||\dot \bk||^2 \ .
\eeq

\subsection{First result: $\bk$ be geodesic}

The third order field equations (\ref{ricci3}) imply that $\bk$ be
geodesic. Then it can be normalized so that $\dot \bk = 0$. The
optical scalars (\ref{om_def}) and (\ref{shear_def}) thus become
\begin{align}
\omega^2&=\frac12 \eta^{\alpha\mu}\eta^{\beta\nu}k_{[\beta , \alpha]}k_{\mu, \nu}\ , \nonumber\\
\theta^2+|\sigma|^2&=\frac12 \eta^{\alpha\mu}\eta^{\beta\nu}k_{(\beta , \alpha)}k_{\mu, \nu}\ .
\end{align}
The second order Ricci tensor (\ref{ricci2}) simplifies to
\beq
\label{secondorder}
    \R<2>_{\aa\bb} = 2 H  \mD k_\aa k_\bb\ , \qquad
    \mD = \ddot H + 2\theta\dot H +4H\omega^2\ ,
\eeq
leading to the condition $\mD =0$, which gives the following
equation for $H$
\beq
\label{eq1H}
0=\ddot H + 2\theta\dot H
+4H\omega^2\ .
\eeq
Finally, the first order Ricci tensor
(\ref{ricci1})--(\ref{ricci1_coeffs}) becomes
\beq\label{Rone}\begin{split}
    \R<1>_{\aa\bb} &=\eta^{\lambda\gamma}H_{,\lambda\gamma} k_\aa k_\bb +
    2k_{(\aa}B_{\bb)}\\
      &\quad\, -2\left[(\dot H + 2\theta H)k_{(\alpha,\beta)}
       -\eta^{\lambda\gamma}Hk_{\alpha , \gamma}k_{\beta , \lambda}\right]\ ,\\
\end{split}\eeq
with
\beq
B_\bb = -(\dot H + 2\theta H)_{,\bb}+\eta^{\lambda\gamma}(2H_{,\lambda}k_{\beta , \gamma}+Hk_{\bb,\gamma\lambda})\ .
\eeq

The vector $\bk$ is an eigenvalue of the Ricci tensor, i.e.
\beq
    R_{\aa\sigma}k^\sigma =(B_\sigma k^\sigma )k_\aa\ .
\eeq

It proves easier to handle with the remaining set of first order field equations by specifying a general field of real null direction in Minkowski space together with an adapted tetrad frame, then setting to zero each individual frame component of the first order Ricci tensor.

\subsection{Simplified tetrad procedure}

Following \cite{debney,kerr2} introduce the set of null
coordinates in Minkowski space $(u,v,\zeta,\bar\zeta)$, which are
related to the standard Cartesian coordinates $(t,x,y,z)$ by
\begin{align}
u&=\frac{1}{\sqrt{2}}(t-z)\ , \qquad v=\frac{1}{\sqrt{2}}(t+z)\ , \nonumber\\
\zeta&=\frac{1}{\sqrt{2}}(x+iy)\ , \,\,\,\quad \bar\zeta=\frac{1}{\sqrt{2}}(x-iy)\ .
\end{align}
The metric (\ref{ansatz}) becomes
\begin{equation}
\rmd s^2 = 2(\rmd\zeta\rmd\bar\zeta-\rmd u\rmd v)-2\ee Hk_\alpha k_\beta\rmd x^\alpha\rmd x^\beta\ .
\end{equation}
A general field of real null directions in Minkowski space is given by
\begin{equation}
\label{k_comp}
k=-[\rmd u+ Y{\bar Y}\rmd v+{\bar Y}\rmd\zeta+Y\rmd\bar\zeta]\ , \qquad
\bk= Y{\bar Y}\partial_u +\partial_v-Y\partial_\zeta-{\bar Y}\partial_{\bar\zeta}\ ,
\end{equation}
where $Y$ is an arbitrary complex function of coordinates. In fact
the independent components of $\bk$ reduce to two real functions
of the coordinates, due to the two conditions 1) $\bk$ forms a
lightlike world line and 2) $\bk$ has an arbitrary
parametrization. In Eq. (\ref{k_comp}) these two real functions of
the coordinates collapsed in a single complex function $Y$, namely
$\bk=\bk(Y,\bar Y)$.

We introduce the following frame
\begin{equation}
\oo^1=\rmd\zeta+Y\rmd v\ , \quad
\oo^2=\rmd\bar\zeta+{\bar Y}\rmd v\ , \quad
\oo^3=-k\ , \quad
\oo^4=\rmd v+\ee H\oo^3\ ,
\end{equation}
so that
\beq
\rmd s^2= 2\oo^1 \oo^2 - 2\oo^3 \oo^4\ .
\eeq
The dual frame is
\begin{equation}
\be_1=\partial_\zeta-{\bar Y}\partial_u\ , \quad
\be_2=\partial_{\bar\zeta}-Y\partial_u\ , \quad
\be_3=\partial_u-\ee H\bk\ , \quad
\be_4=\bk\ .
\end{equation}
The connection coefficients are given by
\beq
\Gamma_{cab}=-e_c{}^{\mu}e_{a\,\mu;\nu}e_b{}^{\nu}\ .
\eeq
Note that $\oo^1_\mu=-k_{\mu,{\bar Y}}$ and $\oo^2_\mu=-k_{\mu,Y}$, trivially implying $\oo^1(\bk)=0=\oo^2(\bk)$, because
\beq
\bk \cdot \oo^1=\eta^{\alpha\beta}k_\alpha \oo^1_\beta=-\eta^{\alpha\beta}k_\alpha k_\beta{}_{,{\bar Y}}=-k^\alpha k_\alpha{}_{,{\bar Y}}= 0\ .
\eeq
Similarly $\bk \cdot \oo^2=0$.

The derivative of $\bk$ is quite simple
\beq
    k_{\mu,\nu} = k_{\mu,{\bar Y}}{\bar Y}_{,\nu}+k_{\mu,{\bar Y}}{\bar Y}_{,\nu}= -\oo^1_\mu\bar Y_{,\nu} - \oo^2_\mu Y_{,\nu}\ .
\eeq

Next introduce the following standard notation for the directional derivatives along the frame vectors
\begin{align}
\label{frameder}
D&\equiv\nabla_{\bk}\,=\partial_v+Y{\bar Y}\partial_u-Y\partial_{\zeta}-{\bar Y}\partial_{\bar\zeta}\ , \nonumber\\
\Delta&\equiv\nabla_{\be_3}=\partial_u-\ee HD\ , \nonumber\\
\delta&\equiv\nabla_{\be_1}=\partial_\zeta-{\bar Y}\partial_u\ .
\end{align}
The geodesic curvature $\kappa$, complex expansion $\rho$ and shear $\sigma$ of the null congruence $\bk$ are given by
\begin{align}
\label{krsdefs}
\kappa&\equiv-\Gamma_{414}=-k^{\alpha}De_{1\alpha}=D{\bar Y}\ , \nonumber\\
\rho&\equiv-\Gamma_{412}=-k^{\alpha}\bar\delta e_{1\alpha}=\bar\delta{\bar Y}\ , \nonumber\\
\sigma&\equiv-\Gamma_{411}=-k^{\alpha}\delta e_{1\alpha}=\delta{\bar Y}\ ,
\end{align}
respectively.
It is also useful to introduce the quantity
\beq
\label{taudef}
\tau\equiv-\Gamma_{413}=-k^{\alpha}\Delta e_{1\alpha}=\partial_u{\bar Y}\ .
\eeq

If the principal null vector $\bk$ is geodesic, then $\kappa=0$, i.e.
\begin{equation}
\label{cond1}
0=D{\bar Y}={\bar Y}_{,v}+Y{\bar Y}{\bar
Y}_{,u}-Y{\bar Y}_{,\zeta}-{\bar Y}{\bar Y}_{,\bar\zeta}\ .
\end{equation}
If it is also shearfree, then $\sigma= 0$, i.e.
\begin{equation}
\label{cond2}
0=\delta {\bar Y}={\bar Y}_{,\zeta}-{\bar Y}{\bar
Y}_{,u}\ , \qquad \to \qquad ({\rm c.c.}) \qquad
0=Y_{,\bar\zeta}-YY_{,u}\ ,
\end{equation}
where ``c.c.'' stands for ``complex conjugate.''
Substituting it into Eq. (\ref{cond1}) then yields
\begin{equation}
\label{cond1new}
0={\bar Y}_{,v}-{\bar Y}{\bar Y}_{,\bar\zeta}\ ,
\qquad \to \qquad ({\rm c.c.}) \qquad 0=Y_{,v}-YY_{,\zeta}\ .
\end{equation}
The conditions (\ref{cond2}) and (\ref{cond1new}) thus give
\beq
    Y_{,\bar\zeta}= YY_{,u}\ ,\qquad Y_{,v} = YY_{,\zeta}\ ,
\eeq
whence if ${\Large \Box}_0$ is the flat-space wave operator, then
\beq
   {\Large \Box}_0 Y \equiv \eta^{\aa\bb}Y_{,\aa\bb} = 2Y_{,\bar\zeta\zeta}-2Y_{,uv} = (Y^2)_{,u\zeta} - (Y^2)_{,\zeta u} = 0\ ,
\eeq
and therefore $Y$ is a solution of the wave equation in Minkowski space whenever the congruence is geodesic and shearfree.
They also show that the congruence $\bk$ must satisfy the Kerr Theorem, i.e. $Y$ is a root of an analytic equation
\begin{equation}
\label{Fimpl}
0=F(Y, {\bar\zeta}Y + u, vY + \zeta)\ ,
\end{equation}
where $F$ is an arbitrary function analytic in the three complex variables $Y$, ${\bar\zeta}Y + u$ and $vY + \zeta$.

\subsection{Completion of the solution}

In terms of the connection coefficients previously introduced the optical scalars write as
\beq
\theta=-\frac12(\rho+\bar\rho)\ , \qquad
\omega^2=-\frac14(\rho-\bar\rho)^2\ ,
\eeq
so that the single equation (\ref{eq1H}) coming from the vanishing of second order Ricci tensor reads
\beq
\label{eq1Hn}
0=\ddot H-(\rho+\bar\rho)\dot H-(\rho-\bar\rho)^2H\ .
\eeq

The nonvanishing relevant frame components of the first order Ricci tensor (\ref{Rone}) are given by
\begin{subequations}
\begin{align}
    \R<1>_{11} &= \ 2\sigma [\dot H - (\bar \rho - \rho)H]\ ,\label{Rone11}\\
    \R<1>_{12} &= (\rho + \bar \rho)\dot H - (\rho^2 +\bar \rho^2- 2\sigma \bar\sigma)H\ ,\label{Rone12}\\
    \R<1>_{13} &= \delta \dot H+(\rho-\bar\rho)\delta H+2\sigma\bar\delta H-\tau \dot H -(\delta\bar\rho+2\bar\tau\sigma+2\tau\rho-\bar\delta\sigma)H\ ,\label{Rone13}\\
    \R<1>_{33} &= 2\left[\delta\bar\delta H-(\rho_{,u}+\bar\rho_{,u})H-\tau\bar\delta H-\bar\tau\delta H-\rho H_{,u}\right]\ ,\label{Rone33}\\
    \R<1>_{34} &= \ddot H-(\rho+\bar\rho)\dot H-(\rho-\bar\rho)^2H\ , \label{Rone34}
\end{align}
\end{subequations}
since $\R<1>_{22}$ and $\R<1>_{23}$ are c.c. of $\R<1>_{11}$ and $\R<1>_{13}$ respectively.
The identities
\begin{align}
\label{spinidentities}
\bar\delta\tau&=\rho_{,u}+\tau\bar\tau\ ,  &\delta\tau=\sigma_{,u}+\tau^2\ , \nonumber\\
\delta\rho&=\bar\delta\sigma+\tau(\rho-\bar\rho)\ ,   &D\rho=\sigma\bar\sigma+\rho^2\ , \nonumber\\
D\tau&=\bar\tau\sigma+\tau\rho\ , &D\sigma=\sigma(\rho+\bar\rho)\
,
\end{align}
as well as the commutation relations
\begin{align}
\label{commutationrels}
\partial_uD-D\partial_u&=-\bar\tau\delta-\tau\bar\delta\ , & \delta D-D\delta=-\bar\rho\delta-\sigma\bar\delta\ , \nonumber\\
\delta\partial_u-\partial_u\delta&=\tau\partial_u\ , & \bar\delta\delta-\delta\bar\delta=-(\rho-\bar\rho)\partial_u\ ,
\end{align}
have been used here to simplify the expressions involving frame derivatives of $H$.
Setting to zero each component of Eqs. (\ref{Rone11})--(\ref{Rone34}) gives a set of first order equations.
Note that the condition coming from Eq. (\ref{Rone34}) is the same as Eq. (\ref{eq1Hn}).

Equation (\ref{Rone11}) implies $\sigma=0$, i.e. the congruence $\bk$ must be shearfree.
The remaining first order equations thus simplify as
\begin{subequations}
\begin{align}
    0&= (\rho + \bar \rho)\dot H - (\rho^2 +\bar \rho^2)H\ ,\label{Rone12n}\\
    0&= \delta \dot H+(\rho-\bar\rho)\delta H-\tau \dot H -(\delta\bar\rho+2\tau\rho)H\ ,\label{Rone13n}\\
    0&=\delta\bar\delta H-(\rho_{,u}+\bar\rho_{,u})H-\tau\bar\delta H-\bar\tau\delta H-\rho H_{,u}\ ,\label{Rone33n}
\end{align}
\end{subequations}
and the identities (\ref{spinidentities}) become
\begin{align}
\label{spinidentitiesnew}
\bar\delta\tau&=\rho_{,u}+\tau\bar\tau\ ,  &\delta\tau=\tau^2\ , \nonumber\\
\delta\rho&=\tau(\rho-\bar\rho)\ ,  &\dot\rho=\rho^2\ , \nonumber\\
\dot\tau&=\tau\rho\ . &
\end{align}
Taking the $\delta$ derivative of Eq. (\ref{Rone12n}) together with Eq. (\ref{Rone13n}) gives rise to the following compatibility condition
\beq
\label{compat1}
\rho(\rho+\bar\rho)\delta H=[\tau\rho(\rho+\bar\rho)+\tau\bar\rho(\rho+3\bar\rho)+\rho\delta\bar\rho]H\ .
\eeq
Take the complex conjugate of this equation and then its $\delta$ derivative; Eq. (\ref{Rone33n}) thus gives rise to a second compatibility condition
\beq
\label{compat2}
(\rho+\bar\rho)H_{,u}=\left[3\left(\frac{\rho\bar\tau}{\bar\rho^2}\delta\bar\rho+\frac{\bar\rho\tau}{\rho^2}\bar\delta\rho\right)+(\bar\rho+3\rho)\frac{\bar\rho_{,u}}{\bar\rho}+(\rho-3\bar\rho)\frac{\rho_{,u}}{\rho}+6\frac{\tau\bar\tau}{\bar\rho}\right]H\ .
\eeq

By using Eq. (\ref{Rone12n}), Eq. (\ref{eq1Hn}) rewrites as
\begin{equation}
\label{eqh2}
\ddot H=2(\rho+\bar\rho)\dot H-2\rho\bar\rho H\ .
\end{equation}
Let the complex expansion be nonzero, i.e. $\rho\not=0$.
It is easy to check that $\rho\bar\rho$ and $\rho+\bar\rho$ are particular solutions, and therefore the general solution is
\begin{equation}
\label{hsolgen}
H=\frac12M(\rho+\bar\rho)+B\rho\bar\rho\ , \qquad
\dot M=\dot B=0\ ,
\end{equation}
where $M(Y,{\bar Y})$ and $B(Y,{\bar Y})$ are real functions of
$Y$ and ${\bar Y}$. Substituting the general solution
(\ref{hsolgen}) for $H$ into Eq. (\ref{Rone12n}) one easily gets
$B=0$, by using the relation $\dot\rho=\rho^2$, so that
\begin{equation}
\label{hsol2}
H=\frac12M(\rho+\bar\rho)\ .
\end{equation}

Substituting now this solution for $H$ into Eq. (\ref{compat1}) leads to
\begin{equation}
\label{compat1new}
\delta M=\frac{3M}{\rho}\tau\bar\rho\ .
\end{equation}
But $M=M(Y, {\bar Y})$, so that $\delta M=M_{,Y}\bar\rho$,
implying that
\beq \label{eqM}
M_{,Y}=\frac{3M}{\rho}\tau\ ,
\qquad M_{,\bar Y}=\frac{3M}{\bar\rho}\bar\tau\ .
\eeq
The second
compatibility condition (\ref{compat2}) then yields
\beq
\label{compat2new}
\frac{\bar\rho}{\rho}\left(\bar\delta\tau-\frac{\tau}{\rho}\bar\delta\rho\right)-
{\rm c.c.}=0\ ,
\eeq
where the relation
\beq
M_{,u}=3M\tau\bar\tau\left(\frac{1}{\rho}+\frac{1}{\bar\rho}\right)\
\eeq
has been used. Equation (\ref{compat2new}) is an additional
equation for $Y$ and ${\bar Y}$ which we will discuss later.

Following the original work \cite{debney} we now introduce
$P=(M/m)^{-1/3}$, where $m$ is a real constant. The first equation
of (\ref{eqM}) thus becomes
\begin{equation}
\label{eqP}
P^{-1}P_{,Y}=-\frac{\tau}{\rho}\ .
\end{equation}
By taking $\delta$ of both sides we then find
\beq
\label{eqP2}
-\bar\rho P^{-2} (P_{,Y})^2+\bar\rho
P^{-1}P_{,YY}=-\frac{\tau^2}{\rho^2}\bar\rho=-\bar\rho P^{-2}
(P_{,Y})^2\ ,
\eeq
since $\delta P=\bar\rho P_{,Y}$ and $\delta
P_{,Y}=\bar\rho P_{,YY}$, and the identities
(\ref{spinidentitiesnew}) have been used to replace $\delta\rho$
and $\delta\tau$ on the RHS. Equation (\ref{eqP2}) thus implies
$P_{,YY}=0$, whose solution is
\beq P=pY{\bar Y}+qY+{\bar q}{\bar
Y}+c\ ,
\eeq
where $p$ and $c$ are real constants and $q$ is a
complex constant.

Let us turn to the remaining compatibility condition
(\ref{compat2new}). First note that it can be equivalently
rewritten as
\beq
\label{compat2new2}
\bar\rho\bar\delta\left(\frac{\tau}{\rho}\right)- {\rm c.c.}=0\ .
\eeq
By using Eq. (\ref{eqP}) we have
\beq
\bar\rho\bar\delta\left(\frac{\tau}{\rho}\right)=\rho\bar\rho
P^{-1}[P^{-1}P_{,Y}P_{,\bar Y}-P_{,Y{\bar Y}}]\ .
\eeq
Take the
complex conjugate of this expression taking into account that $P$
is real; substituting then into Eq. (\ref{compat2new2}) we find
that it is identically satisfied.

Finally, taking the exterior derivative of $Y$ gives
\begin{align}
\label{eqY}
\rmd Y&=\delta Y\omega^1+Y_{,u}\omega^3=P^{-1}\bar\rho[P\omega^1-P_{,\bar Y}\omega^3]\nonumber\\
&=P^{-1}\bar\rho[(qY+c)(\rmd\zeta+Y\rmd v)-(pY+{\bar q})(\rmd
u+Y\rmd\bar\zeta)]\ ,
\end{align}
whose general solution is
\begin{equation}
\label{Fdef}
0=F\equiv\phi(Y)+(qY+c)(\zeta+Yv)-(pY+{\bar
q})(u+Y\bar\zeta)\ ,
\end{equation}
according to Eq. (\ref{Fimpl}), with $\phi$ an arbitrary analytic
function of the complex variable $Y$. In fact, differentiating Eq.
(\ref{Fdef}) leads to
\beq
\label{eqF1}
F_{,Y}\rmd Y=\rmd
F=F_{,\alpha}\rmd x^\alpha=(qY+c)(\rmd\zeta+Y\rmd v)-(pY+{\bar
q})(\rmd u+Y\rmd\bar\zeta)\ .
\eeq
Furthermore, taking the $\delta$ derivative of $F$, i.e.
\beq
\bar\rho F_{,Y}=\delta
F=(\partial_\zeta-{\bar Y}\partial_u)F=P\ ,
\eeq
implies that the complex expansion of the null vector $\bk$ is given by
\beq
\bar\rho=PF_{,Y}{}^{-1}\ .
\eeq
Equation (\ref{eqY}) then immediately follows.

Summarizing, the solution is given by
\begin{equation}
\rmd s^2 = 2(\rmd\zeta\rmd\bar\zeta-\rmd u\rmd v)-\frac{m}{P^3}(\rho+\bar\rho)[\rmd u+ Y{\bar Y}\rmd v+{\bar Y}\rmd\zeta+Y\rmd\bar\zeta]^2\ ,
\end{equation}
with
\beq
P=pY{\bar Y}+qY+{\bar q}{\bar Y}+c\ , \qquad
\bar\rho=PF_{,Y}{}^{-1}\ .
\eeq

The main properties of such a family of solutions are listed below
(see e.g. \cite{ES}):

\begin{itemize}

\item[1.] They are all algebraically special, with $\bk$ shearfree
and geodesic.

\item[2.] They all admit at least a one-parameter group of motions
with Killing vector
\beq
\boldsymbol{\xi}=c\partial_u+p\partial_v+{\bar
q}\partial_\zeta+q\partial_{\bar\zeta}\ ,
\eeq
which is
simultaneously a Killing vector of flat spacetime. The solutions
can be simplified by performing a Lorentz transformation. One can
thus assume that if

\begin{itemize}

\item[a)] $\eta_{\alpha\beta}\xi^\alpha\xi^\beta<0$, then
$P=(1+Y{\bar Y})/\sqrt{2}$, i.e. with $\boldsymbol{\xi}$ pointing
along the $u+v$ (or $t$) direction ($p=c=1/\sqrt{2}$, $q=0$);

\item[b)] $\eta_{\alpha\beta}\xi^\alpha\xi^\beta>0$, then
$P=(1-Y{\bar Y})/\sqrt{2}$, i.e. with $\boldsymbol{\xi}$ pointing
along the $v-u$ (or $z$) direction ($-p=c=1/\sqrt{2}$, $q=0$);

\item[c)] $\eta_{\alpha\beta}\xi^\alpha\xi^\beta=0$, then $P=1$,
i.e. with $\boldsymbol{\xi}$ pointing along the $u$ direction
($p=q=0$, $c=1$).

\end{itemize}

\item[3.] For a timelike Killing vector $\boldsymbol{\xi}$, the
particular case $\phi=-iaY$, with $m=\mathcal M$, leads to the
Kerr solution (\ref{KSmetric1})--(\ref{KSmetric5}), once written
in Kerr-Schild coordinates.

\end{itemize}

\section{Concluding remarks}

We have presented an alternative derivation of Kerr solution by treating Kerr-Schild metrics as {\it exact linear perturbations} of Minkowski spacetime.
In fact they have been introduced as a linear superposition of the flat spacetime metric and a squared null vector field $\bk$ multiplied by a scalar function $H$.

In the case of Kerr solution the vector $\bk$ is geodesic and shearfree and it is independent of the mass parameter ${\mathcal M}$, which enters instead the definition of $H$ linearly.
This linearity property allows one to solve the field equations order by order in powers of $H$ in complete generality, i.e. without any assumption on the null congruence $\bk$.
The Ricci tensor turns out to consist of three different contributions.
Third order equations all imply that $\bk$ must be geodesic; it must be also shearfree as a consequence of first order equations, whereas the solution for $H$ comes from second order equations too.

The present treatment can be generalized to include also the
electromagnetic field, i.e. to the case of Kerr-Newman. In fact,
even in the charged Kerr solution the congruence of $\bk$-lines
depend only on the rotation parameter $a$ and not on the mass
$\mathcal M$ or charge $Q$. Furthermore, the electromagnetic field
is linear in $Q$ and the metric is linear in $\mathcal M$ and
$Q^2$, since the function $H$ is obtained simply by replacing
${\mathcal M}\to{\mathcal M}-Q^2/(2r)$.

\section*{Acknowledgements}

We thank Prof. R. Ruffini and ICRANet for support.

\end{document}